\documentclass[graybox]{svmult}

\usepackage{type1cm}        
\usepackage{makeidx}         
\usepackage{graphicx}        
\usepackage{multicol}        
\usepackage[bottom]{footmisc}

\usepackage{url}

\usepackage{newtxtext}       %
\usepackage{newtxmath}       
\usepackage{comment}
\usepackage{braket}

\usepackage{bbm}
\usepackage{bm}

\newcommand\bbe{\mathbb{E}}

\newcommand\bbp{\mathbb{P}}

\newcommand\bbr{\mathbb{R}}

\newcommand\calE{\mathcal{E}}

\newcommand\calM{\mathcal{M}}

\newcommand\calR{\mathcal{R}}

\newcommand\calX{\mathcal{X}}
\newcommand\calY{\mathcal{Y}}

\makeindex             

\begin{document}

\title*{The interplay of robustness and generalization in quantum machine learning}
\author{Julian Berberich, Tobias Fellner, Christian Holm}
\institute{Julian Berberich \at University of Stuttgart, Institute for Systems Theory and Automatic Control, \email{julian.berberich@ist.uni-stuttgart.de}
\and Tobias Fellner \at University of Stuttgart, Institute for Computational Physics \email{tobias.fellner@icp.uni-stuttgart.de}
\and Christian Holm \at University of Stuttgart, Institute for Computational Physics \email{holm@icp.uni-stuttgart.de}}
%
%
\maketitle

\abstract{
While adversarial robustness and generalization have individually received substantial attention in the recent literature on quantum machine learning, their interplay is much less explored.
In this chapter, we address this interplay for variational quantum models, which were recently proposed as function approximators in supervised learning.
We discuss recent results quantifying both robustness and generalization via Lipschitz bounds, which explicitly depend on model parameters.
Thus, they give rise to a regularization-based training approach for robust and generalizable quantum models, highlighting the importance of trainable data encoding strategies.
The practical implications of the theoretical results are demonstrated with an application to time series analysis.
}


\section{Introduction}

Adversarial attacks are data perturbations which aim to cause failures in machine learning algorithms.
It is well-known in classical machine learning that even adversarial attacks of very small size can lead to misclassification~\cite{goodfellow2014explaining,szegedy2014intriguing}.
Therefore, significant efforts were made to develop robust training methods which improve adversarial robustness and, thereby, the practical reliability of machine learning algorithms~\cite{wong2018provable,tsuzuku2018lipschitz,madry2019towards}.
Beyond its protection against adversarial attacks, the importance of robustness stems from its close connection to generalization.
The generalization performance of a machine learning algorithm refers to its ability of extrapolating from the training data set to (unseen) test data. 
The interplay of robustness and generalization is well-explored in the classical machine learning literature~\cite{goodfellow2014explaining,szegedy2014intriguing,krogh1991simple,xu2012robustness,papernot2016distillation}.
In particular, enhanced robustness typically improves generalization performance, which can be explicitly quantified via generalization bounds, i.e., bounds on the difference between the expected testing loss and the empirical training loss~\cite{xu2012robustness}.

In this chapter, we study the interplay of robustness and generalization in quantum machine learning (QML).
QML is a recently emerging field at the intersection of classical machine learning and quantum computing~\cite{biamonte2017quantum,cerezo2022challenges}.
This chapter focuses on the usage of quantum computers as machine learning models.
Various recent works have highlighted the importance of adversarial robustness in this context, demonstrating the sensitivity of quantum models w.r.t.\ adversarial attacks and proposing training strategies for improved robustness~\cite{edwards2020quantum,liu2020vulnerability,lu2020quantum,du2021quantum,guan2021robustness,liao2021robust,weber2021optimal,gong2022universal,west2023towards,west2023benchmarking,berberich2024training}.
Generalization is another key property of quantum models.
In particular, a substantial body of literature has developed bounds on the generalization performance~\cite{berberich2024training,abbas2021power,banchi2021generalization,caro2021encoding,huang2021power,caro2022generalization,caro2023out,peters2023generalization,jerbi2024shadows}.
However, most of these bounds are uniform and do not involve model parameters, which leads to substantial limitations~\cite{gil2024understanding}.

In this chapter, we focus on a particular QML approach which uses variational quantum algorithms (VQAs) for supervised learning~\cite{benedetti2019parameterized,havlicek2019supervised,schuld2019quantum}.
A VQA combines a parameterized quantum circuit with a classical optimization routine, see~\cite{cerezo2021variational} for an overview.
We discuss the robustness of variational quantum models against adversarial attacks using Lipschitz bounds~\cite{berberich2024training}, which are a powerful tool from classical machine learning and control theory~\cite{fazlyab2019efficient,revay2021convex,pauli2022training}.
Further, we emphasize the strong links between robustness and generalization in QML by stating a generalization bound from~\cite{berberich2024training} which, in contrast to alternative results, depends explicitly on model parameters and, therefore, overcomes their limitations.
We also discuss the implications of these results on the data encoding in QML, highlighting the important benefits of trainable encoding strategies.
With novel numerical results, we demonstrate that the considered robustness perspective not only brings theoretical insights but also relevant practical advantages.

The chapter is structured as follows.
After introducing the considered quantum model class in Section~\ref{sec:vq_models}, we separately address robustness and generalization in Section~\ref{sec:robustness} and Section~\ref{sec:generalization}, respectively.
Section~\ref{sec:numerics} contains our numerical results on time series analysis, and we conclude the chapter in Section~\ref{sec:conclusion}.

\section{Variational quantum models}\label{sec:vq_models}
This chapter focuses on variational quantum models, which rely on parameterized unitary operators
\begin{align}\label{eq:Uj_def}
    U_{j,\Theta_j}(x)=e^{-i(w_j^\top x+\theta_j)H_j}
\end{align}
for $j=1,\dots N$.
Here, $x\in\bbr^d$ denotes the classical data that are considered, e.g., for classification tasks.
Moreover, $\Theta_j=\{w_j,\theta_j\}$ are parameters that are updated during training using an optimization strategy introduced below in more detail.
Finally, the $H_j$'s are Hermitian generators, i.e., they satisfy $H_j=H_j^\dagger$, and they may act on one or multiple qubits.
Based on the operators $U_{j,\Theta_j}(x)$, we define the parameterized quantum circuit
\begin{align}\label{eq:U_def}
    U_{\Theta}(x)=U_{N,\Theta_N}(x)\cdots U_{1,\Theta_1}(x).
\end{align}
For any data sample $x$, a set of trainable parameters $\Theta=\{\Theta_1,\dots,\Theta_N\}$, and Hermitian generators $H_j$, the expression~\eqref{eq:U_def} constitutes a valid quantum circuit that can be implemented on a quantum computer (after an additional compilation step, if necessary).
For example, a common choice for $H_j$ are Pauli operators acting on individual qubits 
\begin{align}
    I\otimes\dots\otimes P\otimes\dots\otimes I
\end{align}
with $P\in\{I,X,Y,Z\}$, in which case $U_{j,\Theta_j}(x)$ is a rotation gate with angle $w_j^\top x+\theta_j$.
The circuit may also include constant (parameter-independent) gates such as entangling CNOT gates, for which $w_j=0$ and $\theta_j=1$.

Finally, to obtain an output of the quantum circuit, we perform a measurement.
In particular, repeatedly running $U_{\Theta}(x)$ for the initial state $\ket{0}$ and measuring w.r.t.\ the observable $\calM$, one can retrieve the expectation value 
\begin{align}\label{eq:f_def}
    f_{\Theta}(x)=\braket{0|U_{\Theta}(x)^\dagger\calM U_{\Theta}(x)|0}.
\end{align}

\begin{figure}[t]
    \centering
    \includegraphics[width=0.8\linewidth]{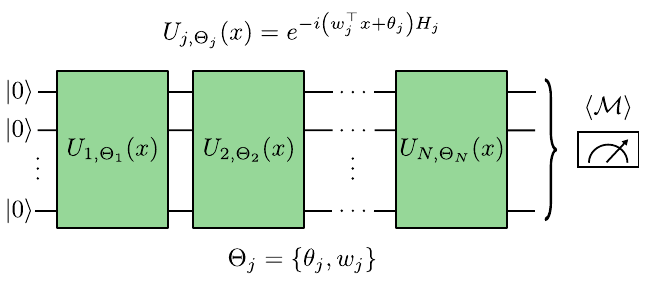}
    \caption{Quantum circuit diagram corresponding to the function $f_{\Theta}(x)$ in Equation~\eqref{eq:f_def}. Unitaries of the form~\eqref{eq:Uj_def} are applied consecutively. The expectation value of an observable $\mathcal{M}$ is measured at the end of the quantum circuit.}
    \label{fig:Circuit_theory}
\end{figure}

For any choice of parameters $\Theta$, $f_{\Theta}$ is a real-valued function $f_{\Theta}:\bbr^d\to\bbr$ whose value can be evaluated using a quantum computer. In Figure~\ref{fig:Circuit_theory}, the corresponding quantum circuit diagram is illustrated. 
In variational QML, this function $f_{\Theta}$ is used as a function approximator to perform learning tasks, e.g., classification or regression.

Note that the considered quantum model includes the full data vector $x$ in each of the $N$ unitaries $U_{j,\Theta_j}(x)$.
This strategy is commonly referred to as data-reuploading and it gives rise to a universal classifier already for a single qubit~\cite{perez2020data}.
In contrast, the majority of existing variational QML approaches employs an alternating structure between data and parameters, i.e., a parameterized quantum circuit of the form $U(\theta)V(x)$ which first encodes the data in $V(x)$ and then acts with a trainable unitary operator $U(\theta)$, where
\begin{align}\label{eq:circuit_alternating}
    U(\theta)=e^{-i\theta_NG_N}\cdots e^{-i\theta_1G_1},\>\>
    V(x)=e^{-ix_1L_1}\cdots e^{-ix_dL_d}
\end{align}
for Hermitian generators $G_j$, $L_j$.
Clearly, the alternating structure $U(\theta)V(x)$ is a special case of $U_{\Theta}(x)$ defined in~\eqref{eq:U_def}, i.e., they are equal for suitable choices of the parameters $\Theta_j$.
Since the encoding of the data $x$ in $V(x)$ is fixed a priori via the choice of $L_j$ and cannot be adapted online, we refer to it as \emph{fixed encoding}.
Conversely, we refer to $U_{\Theta}(x)$ as a circuit with \emph{trainable encoding} since the weights $w_j$ multiplying the data are adapted during training.
It was demonstrated in~\cite{perez2020data,jaderberg2024let} that the additional flexibility provided by a trainable encoding substantially improves the expressivity.
Later in this chapter, we will see that trainable-encoding models also admit crucial advantages over fixed-encoding models in terms of robustness and generalization.

\section{Robustness of quantum models}\label{sec:robustness}

This section focuses on the robustness of the quantum model~\eqref{eq:f_def} using Lipschitz bounds, which are frequently used in classical machine learning and control theory.
In Section~\ref{subsec:robustness_analysis}, we analyze the robustness of a given quantum model by computing a Lipschitz bound involving the model parameters.
Next, in Section~\ref{subsec:robustness_training}, we propose a regularization-based training strategy to train models which are inherently more robust.

\subsection{Robustness analysis}\label{subsec:robustness_analysis}

A machine learning model is referred to as \emph{robust} when small perturbations of the input data do not lead to excessive changes of the output.
Lipschitz bounds are a powerful mathematical tool for characterizing robustness.
For a given quantum model $f_{\Theta}$, we say that $L_{\Theta}>0$ is a Lipschitz bound if\footnote{Throughout this chapter, all involves matrix and vector norms are (induced) $2$-norms, but the results can be adapted to different norms.}
\begin{align}
    \lVert f_{\Theta}(x+\varepsilon)-f_{\Theta}(x)\rVert\leq L_{\Theta}\lVert \varepsilon\rVert
\end{align}
for any $x,\varepsilon\in\bbr^d$.
Intuitively, $L_{\Theta}$ bounds the worst-case change of the function $f_{\Theta}$ when perturbing its input by $\varepsilon$.
Thus, smaller values of $L_{\Theta}$ imply an improved robustness of $f_{\Theta}$.
The following result from~\cite{berberich2024training} provides a Lipschitz bound for the variational quantum model $f_{\Theta}$.

\begin{theorem}\label{thm:lipschitz_bound}
    \cite{berberich2024training} The following is a Lipschitz bound for $f_{\Theta}$
    \begin{align}\label{eq:thm_lipschitz_bound}
        L_{\Theta}=2\lVert\calM\rVert\sum_{j=1}^N\lVert w_j\rVert\lVert H_j\rVert.
    \end{align}
\end{theorem}

The Lipschitz bound~\eqref{eq:thm_lipschitz_bound} depends on the norm of the observable $\calM$, the Hermitian generators $H_j$, and the trainable parameters $w_j$.
In particular, if any of these quantities decreases, then the Lipschitz bound decreases and, therefore, the guaranteed robustness of $f_{\Theta}$ improves.
Lipschitz bounds have found increasing usage in the recent quantum computing literature.
In particular, related Lipschitz bounds were derived for establishing convergence of VQA training~\cite{sweke2020stochastic}, analyzing robustness of quantum algorithms~\cite{berberich2024robustness}, studying quantum models with single shots~\cite{recio2025single}, and investigating double-descent properties of QML~\cite{kempkes2025double}.

While the quantities $\calM$ and $H_j$ appearing in the Lipschitz bound~\eqref{eq:thm_lipschitz_bound} are fixed a priori, the parameters $w_j$ are adapted during training.
As we show in Section~\ref{subsec:robustness_training}, this allows us to systematically influence the robustness of the quantum model during training by penalizing the norm of $w_j$.

\subsection{Training robust quantum models}\label{subsec:robustness_training}

In the following, we employ Theorem~\ref{thm:lipschitz_bound} to derive a robust training strategy for variational quantum models.
To define the training problem, suppose we are given a set of data points $\{x_k,y_k\}_{k=1}^n$ from a domain $\calX\times\calY$.
Based on the data, we want to fit a quantum model $f_{\Theta}$ by minimizing a loss function $\ell$, i.e., by solving
\begin{align}\label{eq:training_problem}
    \min_{\Theta}\frac{1}{n}\sum_{k=1}^n\ell(f_{\Theta}(x_k),y_k).
\end{align}
Theorem~\ref{thm:lipschitz_bound} shows that the norm of the weights $w_j$ influences the Lipschitz bound and, thereby, the robustness of the quantum model.
With this motivation, instead of the original training problem~\eqref{eq:training_problem}, it was proposed by~\cite{berberich2024training} to solve the modified problem 
\begin{align}\label{eq:training_problem_regularized}
    \min_{\Theta}\frac{1}{n}\sum_{k=1}^n\ell(f_{\Theta}(x_k),y_k)+\lambda\sum_{j=1}^N\lVert w_j\rVert^2\lVert H_j\rVert^2.
\end{align}

\begin{figure}[t]
    \centering
    \includegraphics[width=\linewidth]{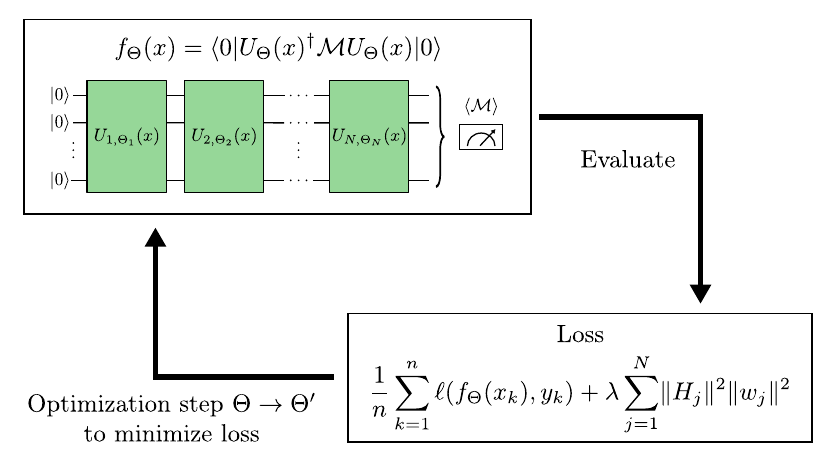}
    \caption{Training scheme of the VQA. The function $f_\Theta(x)$ is evaluated by executing the corresponding quantum circuit. The resulting output, along with the norm of the encoding weights $\lVert \mathbf{w}_j \rVert$, is used to compute a regularized loss function. This loss is minimized iteratively by optimizing all trainable parameters $\Theta$.}
    \label{fig:VQC_Training}
\end{figure}

The variable $\lambda>0$ can be used to trade off the two competing objectives of robustness and training error, and it is a hyperparameter that can be tuned via cross-validation.
The added regularization encourages robustness of the trained quantum model, and it is analogous to ridge regression in classical machine learning~\cite{mohri2018foundations}. Figure~\ref{fig:VQC_Training} shows the iterative update algorithm of the VQA.

Note that the Lipschitz bound in Theorem~\ref{thm:lipschitz_bound} is independent of the parameters $\theta_j$.
For fixed-encoding quantum models, the parameters $\theta_j$ are the only trainable parts of the model and, therefore, the Lipschitz bound~\eqref{eq:thm_lipschitz_bound} is fully determined by offline design choices and cannot be modified during training.
In particular, adding a regularization of the parameters $\theta_j$ to the cost of~\eqref{eq:training_problem_regularized} does not have a direct effect on robustness and, hence, on generalization, compare Section~\ref{sec:generalization}.
Indeed, it was observed in the literature that fixed-encoding models are not overly sensitive w.r.t.\ perturbed data~\cite{mitarai2018quantum,schuld2020circuit}.
However, this also means that robustness cannot be influenced systematically in quantum models with fixed encoding, which leads to important practical limitations, see Section~\ref{sec:numerics} for details.

\section{Generalization of quantum models}\label{sec:generalization}

\begin{figure}[t]
    \centering
    \includegraphics[width=0.8\linewidth]{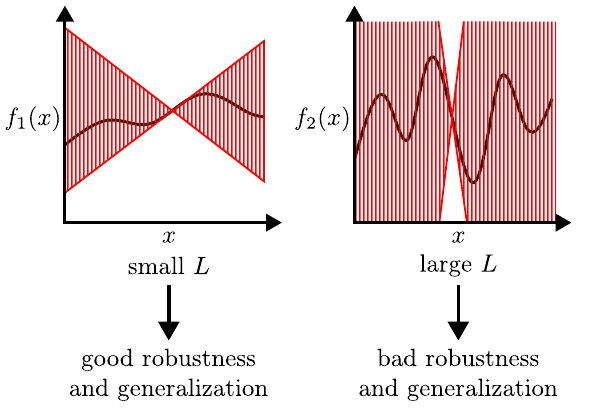}
    \caption{The function on the left-hand side ($f_1(x)$) has a smaller Lipschitz bound than the function on the right-hand side ($f_2(x)$).
    This means that incremental changes from any point to left or right are confined to a smaller range of possible values, indicated via the conic regions shaded in red.
    As a result, the function $f_1(x)$ is smoother, whereas the function $f_2(x)$ can be interpreted as the result of overfitting and can, therefore, be expected to have worse generalization performance.}
    \label{fig:Lischitz_Illustration}
\end{figure}

It is desirable to obtain a quantum model that not only performs well on the training data (i.e., minimizes the training loss in~\eqref{eq:training_problem}), but that also leads to a small loss on test data samples that were not included in the training.
In the following, we show that the Lipschitz bound~\eqref{eq:thm_lipschitz_bound} can be used to quantify the ability of the quantum model~\eqref{eq:f_def} to generalize.
To state the formal result, let us assume that the data $(x,y)$ are sampled according to some distribution $\bbp$.
The central goal of the considered supervised learning task can be formulated as follows:
Find a set of parameters $\Theta$ that minimize the \emph{expected risk} over all possible data samples, i.e.,
\begin{align}
    \calR_{\mathrm{exp}}(\Theta)=\bbe_{(x,y)\sim\bbp}\left[\ell(f_{\Theta}(x),y)
    \right].
\end{align}
However, $\calR_{\mathrm{exp}}(\Theta)$ cannot be evaluated since the distribution $\bbp$ is typically unknown and we only have access to the training data $\{x_k,y_k\}_{k=1}^n$.
Hence, we use the data to define a proxy for the expected risk: the \emph{empirical risk}.
The latter is defined as the cost in the training problem~\eqref{eq:training_problem}, i.e.,
\begin{align}
    \calR_{\mathrm{emp}}(\Theta)=\frac{1}{n}\sum_{k=1}^n\ell(f_{\Theta}(x_k),y_k).
\end{align}
The generalization performance of a model describes how well insights obtained from the training data transfer to the unseen test data.
Mathematically, the ability of a model to generalize can be quantified via the \emph{generalization error} $\calE(\Theta)$, which bounds the difference between the expected risk and the empirical risk
\begin{align}\label{eq:generalization_error}
    \calE(\Theta)=|\calR_{\mathrm{exp}}(\Theta)-\calR_{\mathrm{emp}}(\Theta)|.
\end{align}
Intuitively, one can expect that a small Lipschitz bound implies good generalization:
When small data perturbations lead to small output changes, then the model is less affected by overfitting and the insights from finitely many data points in $\calR_{\mathrm{emp}}(\Theta)$ can be more easily transferred to the (infinitely many) data points corresponding to $\calR_{\mathrm{exp}}(\Theta)$.
This is demonstrated with two simple scalar functions in Figure~\ref{fig:Lischitz_Illustration}.

The following result\footnote{We only provide the informal version from~\cite[Theorem IV.1]{berberich2024training} and refer to~\cite[Theorem B.1]{berberich2024training} for the detailed technical statement.} shown in~\cite{berberich2024training} formalizes this intuition by bounding the generalization error $\calE(\Theta)$ in terms of the Lipschitz bound~\eqref{eq:thm_lipschitz_bound}.

\begin{theorem}\label{thm:generalization}
    \cite{berberich2024training} (Informal).
    The generalization error~\eqref{eq:generalization_error} is bounded as 
    \begin{align}\label{eq:thm_generalization}
        \calE(\Theta)\leq C_1\lVert\calM\rVert\sum_{j=1}^N\lVert w_j\rVert\lVert H_j\rVert+\frac{C_2}{\sqrt{n}}
    \end{align}
    for some $C_1,C_2>0$.
\end{theorem}

The generalization error bound~\eqref{eq:thm_generalization} depends on the Lipschitz bound~\eqref{eq:thm_lipschitz_bound} as well as on the data length $n$.
This shows that a small Lipschitz bound as encouraged via the robust training problem~\eqref{eq:training_problem_regularized} not only improves robustness but also generalization.
One can show that the bound~\eqref{eq:thm_generalization} vanishes in the infinite data limit $n\to\infty$, see~\cite[Appendix B]{berberich2024training} for details.
Further, the generalization error bound is minimized and the best generalization is achieved for $w_j=0$, $j=1,\dots,N$, i.e., when the Lipschitz bound is zero.
Of course, this does not imply that the expected risk is satisfactory since a reduced Lipschitz bound can increase the empirical risk, which may outweigh the reduced generalization error.
In practice, there typically is a sweet spot for which the expected risk can be reduced via the Lipschitz bound by substantially decreasing the generalization error while only mildly increasing the empirical risk (see Section~\ref{sec:numerics} below).

Note that the bound~\eqref{eq:thm_generalization} explicitly depends on the parameters $\Theta$ of the model~\eqref{eq:f_def}.
In contrast, the majority of the existing QML generalization bounds in the literature, e.g.,~\cite{banchi2021generalization,caro2021encoding,huang2021power,caro2022generalization,caro2023out,peters2023generalization,jerbi2024shadows} hold uniformly over all parameters in the model class, with the exception of~\cite{abbas2021power} that relies on the  effective dimension of a quantum model.
The parameter-dependent generalization bound in Theorem~\ref{thm:generalization} has the advantage of leading to targeted training strategies for improved generalization, e.g., via regularization as in~\eqref{eq:training_problem_regularized}.
A more detailed analysis of the limitations of uniform generalization bounds is provided in~\cite{gil2024understanding}.

Finally, we note that, like the Lipschitz bound in Theorem~\ref{thm:lipschitz_bound}, also the generalization bound~\eqref{eq:thm_generalization} is independent of the parameters $\theta_j$.
As a result, for fixed-encoding models, it cannot be influenced systematically during training, which severely limits the possibilities for controlling overfitting (compare Section~\ref{sec:numerics} for the practical implications).

\section{Numerical results}\label{sec:numerics}

To validate our theoretical findings, we train variational quantum machine learning models with the proposed regularization strategy and assess their robustness and generalization capabilities. We begin by motivating and outlining the learning problem used to verify the theoretical statements from Sections~\ref{sec:robustness} and~\ref{sec:generalization} (Section~\ref{subsec:numerics_problem}), followed by a detailed description of the quantum model and training procedure (Section~\ref{subsec:numerics_model}). Finally, we present results on robustness and generalization in Sections~\ref{subsec:numerics_robustness} and~\ref{subsec:numerics_generalization}, respectively. The code to reproduce the numerical results of our study is publicly available on GitHub~\cite{fellner_robustnessgeneralizationqml_2025}.

\subsection{Problem statement}\label{subsec:numerics_problem}
Extracting informative features from time series data is a central challenge across numerous disciplines, including medicine, meteorology, and finance.
Given the critical role of sequential data in many real-world applications, ensuring that models designed for feature extraction are robust to input perturbations is essential.

\begin{figure}[t]
    \centering
    \includegraphics[width=0.8\linewidth]{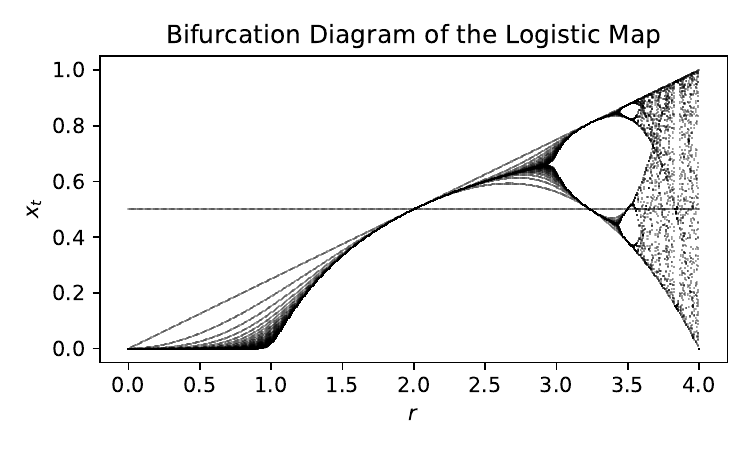}
    \caption{Bifurcation diagram of the logistic map. The first 50 iterations of the logistic map are plotted as a function of the control parameter $r$. For $0 \leq r \leq 3$, the system converges to a single fixed point. In the range $3 \leq r \lesssim 3.5$, the dynamics become periodic, cycling through a finite set of values. For $3.5 \lesssim r \leq 4$, the system enters a chaotic regime, characterized by aperiodicity and sensitive dependence on initial conditions.}
    \label{fig:bifurcation}
\end{figure}

We consider a learning task based on the logistic map~\cite{may_simple_1976}, defined by
\begin{align}\label{eq:logistic_map}
    x_t = r x_{t-1}(1 - x_{t-1})\,,
\end{align}
where the parameter $r \in \mathbb{R}$ governs the dynamics of the system. For $r \in [0, 4]$, the resulting time series remain bounded within the interval $[0,1]$. The behavior of the system varies significantly with $r$: for $0 \leq r \leq 3$, the system converges to a fixed point; for $3 \leq r \lesssim 3.5$, the dynamics become periodic, oscillating between a finite set of values; and for $3.5 \lesssim r \leq 4$, the system exhibits chaotic behavior, characterized by extreme sensitivity to initial conditions~\cite{devaney_introduction_2018}. These different behaviors are visualized in the bifurcation diagram in Figure~\ref{fig:bifurcation}, where we plot the first 50 elements of the sequence $\{x_i\}$ for various values of $r$. Throughout this study, we fix the initial condition to $x_1 = 0.5$.

Chaotic dynamics are of particular importance in modeling complex phenomena such as climate systems and financial markets~\cite{mihailovic_climate_2014, lebaron_chaos_1994}. Given the practical relevance of learning from chaotic time series, we here consider the task of inferring the system parameter $r \in [3.5, 4]$ from a finite sequence $\{x_i\}_{1}^{l}$ generated according to Equation~\eqref{eq:logistic_map}. Formally, the variational quantum machine learning model is trained to approximate the function
\begin{align}
    f_\Theta(\{x_i\}_{1}^{l}) = r\,.
\end{align}

We generate a dataset comprising 1000 sequences, each constructed according to Equation~\eqref{eq:logistic_map}, with values of the parameter $r$ chosen equidistantly from the interval $[3.5, 4]$. We fix the sequence length to $l = 12$. From the resulting dataset of tuples $(\{x_i\}_{1}^{l}, r)$, we randomly allocate 200 samples for training and reserve the remaining 800 for testing purposes.

\subsection{Model and training procedure}\label{subsec:numerics_model}
\begin{figure}[t]
    \centering
    \includegraphics[width=\linewidth]{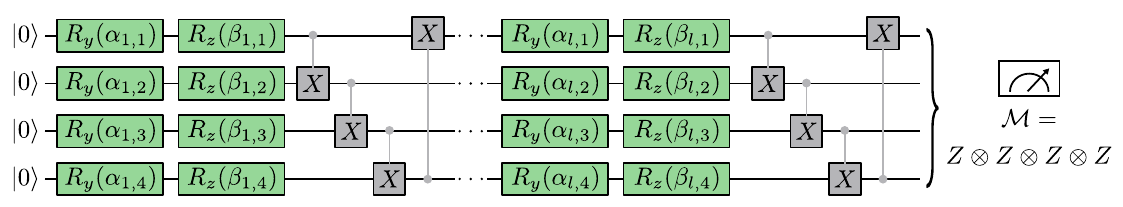}
    \caption{Circuit diagram of the variational quantum circuit employed in this study. The input data sequence is sequentially encoded using Pauli-$Z$ and Pauli-$Y$ rotation gates with angles $\alpha_{i,j}$ and $\beta_{i,j}$, respectively, as defined in Equation~\ref{eq:rotation_angles}. Following each encoding block, controlled Pauli-$X$ (CNOT) gates are applied in a nearest-neighbor pattern to introduce entanglement between adjacent qubits. At the end of the circuit, the observable $\mathcal{M} = Z \otimes Z \otimes Z \otimes Z$ is measured to obtain the output expectation value.}
    \label{fig:quantum_circuit}
\end{figure}

The variational quantum model is designed such that the time series is processed in a sequential manner. The parameterized quantum circuit used in this work is depicted in Figure~\ref{fig:quantum_circuit}. Its design is inspired by a quantum model developed for benchmarking variational quantum machine learning methods on time series prediction tasks~\cite{fellner_quantum_2025a}.

Each data point $x_i$ from the time series is encoded into the quantum circuit via Pauli-$Z$ and Pauli-$Y$ rotation gates, with rotation angles defined as
\begin{align}\label{eq:rotation_angles}
    \alpha_{i,j} &= w_{i,j}^{(1)} x_i + \theta_{i,j}^{(1)}\,, \\
    \beta_{i,j}  &= w_{i,j}^{(2)} x_i + \theta_{i,j}^{(2)}\,,
\end{align}
where $w_{i,j}^{(k)}$ and $\theta_{i,j}^{(k)}$ are trainable parameters, and the subscripts $i$ and $j$ index the time step and the qubit, respectively. These rotations are applied on qubit $j$ to embed the input feature $x_i$. To introduce entanglement between the qubits, we apply a layer of controlled Pauli-$X$ (CNOT) gates in a nearest-neighbor configuration following each data-encoding layer. For our experiments, the quantum circuit operates on four qubits. Finally, at the end of the circuit, we perform a measurement of the multi-qubit observable $\mathcal{M} = Z \otimes Z \otimes Z \otimes Z$ to determine the output of the quantum model. Since the output domain of the quantum model is $[-1,1]$ (the range of possible expectation values when measuring $\mathcal{M}$), we linearly scale the output to the target domain $[3.5,4]$.

The models are simulated using the Python-based quantum machine learning library \texttt{PennyLane}~\cite{bergholm2018pennylane}. Training is carried out using the Adam optimizer~\cite{kingma2014adam} with a learning rate of $0.01$ and proceeds for a total of 2000 epochs.

\subsection{Evaluating model robustness under data perturbations}\label{subsec:numerics_robustness}

\begin{figure}[t]
    \centering
    \includegraphics[width=\linewidth]{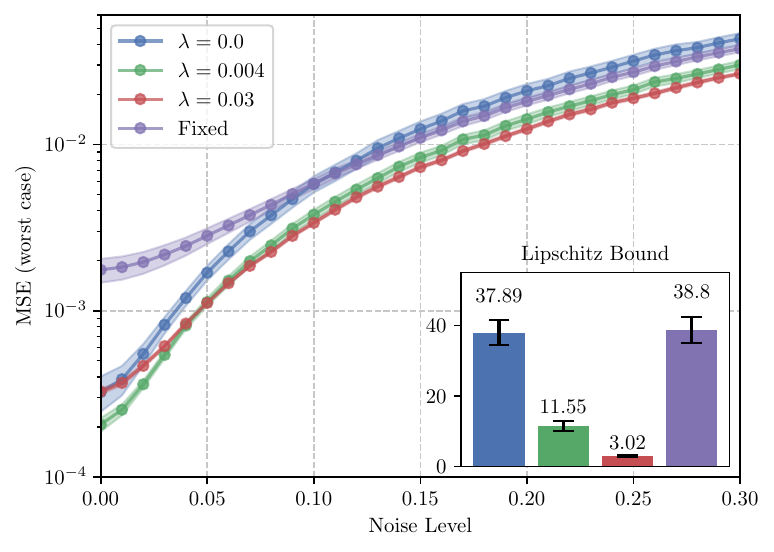}
    \caption{The plot shows the prediction accuracy on the test set as a function of the noise level $\epsilon$ for various trained models. For each data point, 50 models are trained with different random weight initializations. Each model is evaluated on 100 noise-perturbed versions of the test sequences, where the noise is sampled uniformly from the box $[-\epsilon, \epsilon]^l$. From these, the worst-case accuracy for each model is recorded. The plotted values represent the average worst-case accuracy across the 50 trained models, with the shaded regions indicating the standard deviation. We compare four training scenarios: three with trainable encoding using different regularization strengths $\lambda$, and one with a fixed encoding. The inset shows the Lipschitz bounds of the 
    trained models.}
    \label{fig:plot_robustness}
\end{figure}

To empirically validate the theoretical findings on robustness, we train three quantum models using different values of the regularization parameter, $\lambda \in \{0, 0.004, 0.03\}$. In addition, we include a variant with a fixed encoding, in which the parameters $w_{i,j}^{(k)}$ are not optimized but instead held constant throughout training. All trainable parameters are initialized by sampling from a uniform distribution over the interval $[-\pi/2, \pi/2]$.

To evaluate the robustness of the models against input perturbations, we add random noise to the test sequences. Specifically, noise is sampled uniformly from $[-\epsilon, \epsilon]^l$ and added to each input sequence, where $\epsilon$ denotes the noise level. For each value of $\epsilon$, we compute the worst-case mean squared error (MSE) between the predicted and true values of the parameter $r$ over 100 independently perturbed versions of the test set. The results of these experiments are presented in Figure~\ref{fig:plot_robustness}, illustrating how different regularization strategies influence the robustness of the quantum models under noisy input conditions.

As anticipated, the prediction error increases across all models as the noise level $\epsilon$ rises. In the high-noise regime, models trained with a larger regularization parameter $\lambda$ consistently exhibit improved performance. This observation aligns with the theoretical expectations derived from the Lipschitz bound $L_\Theta$.

Specifically, increasing the regularization strength imposes a stronger penalty on the magnitude of the encoding weights during training. As indicated by Equation~\eqref{eq:thm_lipschitz_bound}, smaller encoding weights correspond to a reduced Lipschitz constant. A smaller Lipschitz bound, in turn, implies increased robustness to input perturbations, thereby explaining the enhanced predictive stability of more heavily regularized models under noisy conditions.

The effect of regularization on the Lipschitz bound is illustrated in the inset of Figure~\ref{fig:plot_robustness}, which shows a decrease in $L_\Theta$ for models with trainable encodings as $\lambda$ increases. In contrast, the model with fixed encoding parameters cannot reduce the norm of the encoding weights through optimization. Consequently, its Lipschitz bound remains comparatively large, which contributes to its inferior robustness performance in the presence of noise.

Notably, for small noise levels, there exists a significant performance gap between the model with fixed encoding and the models with trainable encoding. This discrepancy can be attributed to the fact that the model with fixed encoding lacks sufficient expressivity (the ability to approximate a function from the encoded data) due to the reduced number of trainable parameters. Consequently, models with trainable encodings are better equipped to approximate the training data, which translates into improved accuracy on the test set. However, as the noise level increases, the model with fixed encoding demonstrates superior predictive accuracy compared to the model with trainable encoding and regularization parameter $\lambda = 0$. This can be explained by the increased expressivity of the latter model, which, despite its enhanced capacity to fit the training data, may also be more susceptible to overfitting. As a result, it exhibits poorer generalization performance and reduced robustness against data perturbations.

The connection between robustness and generalization also becomes evident when comparing models with trainable encoding in the low-noise regime. Among these, the model trained with $\lambda = 0.004$ achieves the highest predictive accuracy, whereas models with $\lambda = 0$ and $\lambda = 0.03$ perform less accurately. This observation suggests that an intermediate level of regularization can provide an optimal trade-off between expressivity and generalization. In the following section, we will examine generalization performance in more detail.

\subsection{Evaluating model generalization}\label{subsec:numerics_generalization}
\begin{figure}[t]
    \centering
    \includegraphics[width=\linewidth]{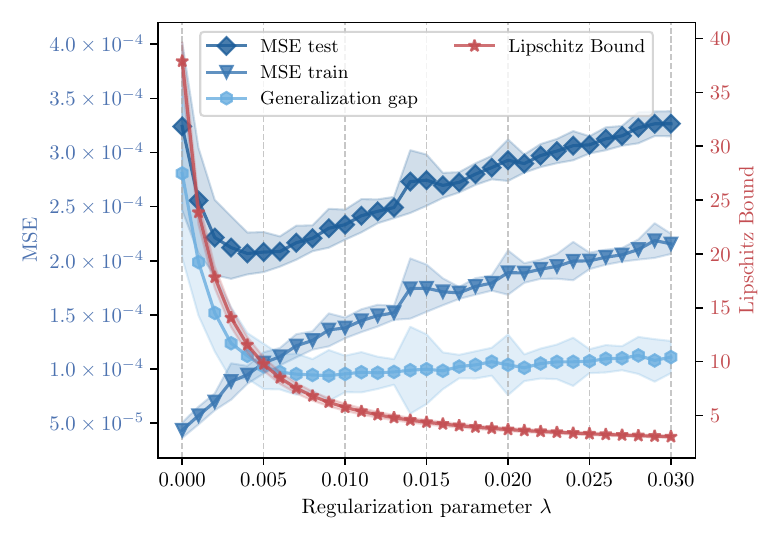}
    \caption{Evaluation of model generalization for varying regularization strengths $\lambda \in [0, 0.03]$. The blue plots show the mean squared error (MSE) on the training and test datasets, as well as the generalization gap. These are affiliated with the left $y$-axis. The Lipschitz bound is displayed in red and is associated with the right $y$-axis. All displayed values are the mean over 50 random weight initializations. The shaded areas correspond to the standard deviation.} 
    \label{fig:plot_generalizatiom}
\end{figure}

To study generalization, we train a series of models over a range of regularization parameters $\lambda \in [0, 0.03]$. The corresponding results are shown in Figure~\ref{fig:plot_generalizatiom}, where we examine how variations in $\lambda$ affect both the prediction accuracy and the Lipschitz bound.

We begin by analyzing the prediction accuracy on the training dataset. As $\lambda$ increases, accuracy steadily declines, which can be attributed to reduced model expressivity resulting from smaller encoding weights. In contrast, on the test dataset, we observe that the prediction error reaches a minimum at $\lambda=0.004$. Relative to the case without regularization this corresponds to an improvement in performance of approximately $50\%$. This suggests that moderate regularization enhances the model’s ability to generalize from the training data to unseen data. However, for larger values of $\lambda$, performance deteriorates again, as the model's expressivity is overly constrained, leading to reduced accuracy on both the training and test datasets.

To gain deeper insight into the model's generalization capability, we examine the generalization gap defined as
\begin{align}
    \mathrm{MSE}_{\text{test}} - \mathrm{MSE}_{\text{train}}\,.
\end{align}
The generalization gap decreases rapidly for small values of $\lambda$, and then stabilizes at an approximately constant level for larger values. This behavior indicates that the regularization approach described in Equation~\eqref{eq:training_problem_regularized} effectively improves generalization. 

In addition, we observe that the Lipschitz bound decreases monotonically with increasing $\lambda$, consistent with the original motivation for incorporating the regularization term, namely, to constrain the model’s Lipschitz constant. Taken together, these findings support the theoretical connection between generalization and the Lipschitz bound as stated in Theorem~\ref{thm:generalization}.

\begin{figure}[t]
    \centering
    \includegraphics[width=\linewidth]{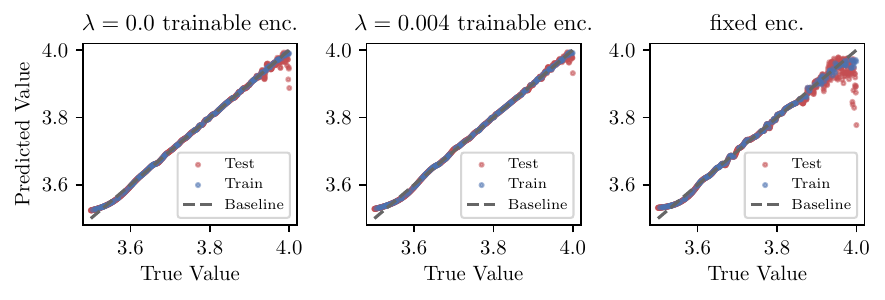}
    \caption{The figure shows the predicted versus true values of $r$ for the training (blue) and test (red) data. The black dashed line indicates perfect prediction. Left and center: Models with trainable encoding without ($\lambda = 0$) and with regularization ($\lambda = 0.004$), respectively. Right: Model with fixed (non-trainable) encoding parameters. The results shown correspond to the median mean squared error on the test data of all 50 random weight initializations.}
    \label{fig:predictions_results}
\end{figure}

To illustrate the generalization performance, we plot the predicted versus true values $r$ for both training and test data in Figure~\ref{fig:predictions_results}. For the trainable encoding with $\lambda=0$ and $\lambda=0.004$, the training points are well fitted. On the test set, performance improves noticeably near $r=4$ when using regularization with $\lambda=0.004$. In contrast, fixing the encoding parameters leads to less accurate predictions, consistent with the high MSE observed for this model in the low-noise regime, see Figure~\ref{fig:plot_robustness}. Notably, all models struggle to predict values near $r=3.5$ and $r=4$, likely due to edge effects in measuring the expectation value $\langle\mathcal{M}\rangle \in [-1,1]$, which is linearly mapped to $[3.5, 4]$. The set of quantum states yielding a given expectation value is largest around $\langle\mathcal{M}\rangle=0$ and shrinks toward the extremes $\langle\mathcal{M}\rangle = \pm1$, making it harder for the model to learn states producing such extreme values.

\subsection{Related numerical results from the literature}

We note that the literature contains further numerical results demonstrating the practicality of Lipschitz bounds in variational QML.
In particular, in~\cite{berberich2024training}, the results shown in Sections~\ref{sec:robustness} and~\ref{sec:generalization} are applied in supervised learning for a classification problem.
Moreover,~\cite{wendlinger2024comparative} compares the robustness of classical models and quantum models, where the latter are analyzed via Theorem~\ref{thm:lipschitz_bound} and trained via~\eqref{eq:training_problem_regularized}.
Finally,~\cite{meyer2024robustness} studies robustness and generalization properties of variational quantum models as in~\eqref{eq:f_def} in the context of quantum reinforcement learning.
The results shown in the present section constitute the first application of Lipschitz-based regularization in QML for time-series analysis.

\section{Conclusion}\label{sec:conclusion}

In this chapter, we reviewed the interplay of robustness and generalization in variational QML.
We introduced theoretical robustness and generalization results from~\cite{berberich2024training} based on Lipschitz bounds.
These results can be used to design regularization strategies for robust training of quantum models.
We also highlighted the role of trainability in the data encoding.
With a numerical application to a time series problem, we demonstrated the practical impact of the presented theoretical results.

While this chapter focuses on the use of Lipschitz bounds and regularization, we note that alternative approaches for handling adversarial robustness in QML exist.
For example, the robustness of a quantum model can be improved during training by including adversarial examples into the training data, see~\cite{west2023towards} for details and references.
Further approaches for enhancing robustness during training include adding quantum noise~\cite{du2021quantum} or solving a min-max optimization problem~\cite{lu2020quantum}, for which tailored generalization bounds can be derived~\cite{georgiou2024adversarial,georgiou2025generalization}.
The key benefit of the robust regularization-based training strategy~\eqref{eq:training_problem_regularized} is that it relies on a rigorous theoretical foundation (Theorem~\ref{thm:lipschitz_bound}) and does not increase the computational complexity.
Studying connections between the robust training approaches from the literature and the presented Lipschitz-based strategy is a promising direction for future research, potentially improving the understanding of robustness and generalization in QML.

\bibliographystyle{IEEEtran}
\bibliography{Literature}  

\end{document}